\documentclass[twocolumn,prb,aps,longbibliography,nobalancelastpage]{revtex4-2}
\usepackage{lineno}
\usepackage{amsmath,amsfonts,amssymb,graphicx,hyperref,epstopdf,xcolor,tikz,scalerel,natbib,array,gensymb}
\usepackage{mathptmx,textcomp,braket, physics}
\usepackage[T1]{fontenc}
\usepackage[utf8]{inputenc}
\usepackage{multirow}
\usepackage{array}
\usepackage{soul}
\DeclareMathAlphabet{\mathpcal}{OMS}{zplm}{m}{n}

\definecolor{lime}{HTML}{A6CE39}
\DeclareRobustCommand{\orcidicon}
{
	\begin{tikzpicture} 
	\draw[lime, fill=lime] (0,0) circle [radius=0.15] node[white] {{\fontfamily{qag}\selectfont \tiny ID}};
	\draw[white, fill=white] (-0.0625,0.095) 	circle [radius=0.007];
	\end{tikzpicture}
	\hspace{-2.2mm}
}
\newcommand\orcidID[1]{\href{https://orcid.org/#1}{\orcidicon}} 
\usepackage{xcolor} 

\newcommand{\be}{\begin {equation}}
\newcommand{\ee}{\end {equation}}
\newcommand{\beqa}{\begin {eqnarray}}
\newcommand{\eeqa}{\end {eqnarray}}

\newcommand{\Exp}[1]{\text{e}^{#1}}

\hypersetup{colorlinks,citecolor=blue,filecolor=black,linkcolor=blue,urlcolor=blue}
 
\begin{document}
\title{Engineering harmonic emission through spatial modulation in a Kitaev chain}

\author{Nivash R \orcidID{0009-0004-3076-9192}}
\email[E-mail: ]{nivash1807@gmail.com}
\author{S. Srinidhi \orcidID{0009-0004-3387-4908}}
\email[E-mail: ]{s.srinidhi312@gmail.com}
\thanks{\\ These authors contributed equally to this work.}
\author{Jayendra N. Bandyopadhyay\orcidID{0000-0002-0825-9370}}
\author{Amol R. Holkundkar\orcidID{0000-0003-3889-0910}}

\affiliation{Department of Physics, Birla Institute of Technology and Science Pilani, Pilani Campus, Vidya Vihar, Pilani, Rajasthan 
333031, India.}

\begin{abstract}
We investigate High-harmonic generation (HHG) in a dimerized Kitaev chain. The dimerization in the model is introduced through a site-dependent modulating potential, determined by a parameter $\lambda \in [-1:1]$. This parameter also determines the strength of the hopping amplitudes and tunes the system's topology. Depending upon the parameter $\lambda$, the HHG emission spectrum can be classified into three segments. The first segment exhibits two plateau structures, with the dominant one resulting from transitions to the chiral partner state, consistent with quasiparticle behavior in the topological superconducting phase. The second segment displays multiple plateaus, where intermediate states enable various transition pathways to higher conduction bands. Finally, the third segment presents broader plateaus, indicative of active interband transitions. In the $\lambda\leq0$ regime, we observe the mid-gap states (MGSs) hybridize with the bulk, suppressing the earlier observed harmonic enhancements. This highlights the key role of the intermediate states, particularly when MGSs are isolated. These results demonstrate that harmonic emission profiles can be selectively controlled through the modulating parameter $\lambda$, offering new prospects for tailoring HHG in topological systems.
\end{abstract}

\maketitle

\section{Introduction}

High-harmonic generation (HHG) is a nonlinear optical process driven by strong light-matter interactions. Traditionally, HHG has been a key technique for generating attosecond pulses and probing electron dynamics in atomic and molecular systems \cite{Corkum2007,RevModPhys}. This decade has opened up new opportunities for researchers to turn their attention to HHG in solids, driven by its potential for technological applications \cite{Ghimire2011,Garg2016,Geneaux2019,Lakhotia2020}. Solid-state HHG is now extending its scope to contemporary frontiers in attosecond science, offering a compact source of extreme ultraviolet (XUV) radiation and pushing the boundaries of attosecond spectroscopy \cite{Hohenleutner2015,You2017,Klemke2019,Bionta2021,Heide2024,Nivash_23,Nivash_24}. HHG in solid-state systems has revealed exciting possibilities, such as observing Bloch oscillations \cite{Schubert2014}, reconstructing band structures \cite{Vampa_15}, generating petahertz currents in solids \cite{petahertz_16}, and investigating Berry curvature effects \cite{berry_18}. Notably, HHG has been demonstrated in materials like monolayer and bilayer graphene \cite{Yoshikawa_17,Mrudul_21}, as well as transition metal dichalcogenides \cite{Yue22}, where it has shed light on the transition dipole moments in two-dimensional materials \cite{TDM_21}. Researchers have recently begun exploring strong-field phenomena in topological materials, which are known for their robustness to external perturbations \cite{Robustness2019,Bauer2019,Bai2021,Baykusheva2021}. HHG in these materials serves as an all-optical probe, revealing unique signatures of topology in the HHG spectra \cite{Silva2019,Alexis_20,Alexis2021,Schmid2021,Heide2022,Bera_23,Neufeld_23,Nivash_AAH,Cabrera2024}. 
Unconventional superconductors are the materials that host topological superconducting phases characterized by non-zero topological invariants \cite{Gurevich2014, Sigrist_DKC}. These systems support exotic quasiparticle excitations, localized at the chain's end, such as Majorana-bound states (MBS). These states are of great interest for fault-tolerant quantum computation due to their non-Abelian statistics \cite{Sarma2015-av,Flensberg2021-ae}. 

To this end, the Kitaev chain model is a paradigmatic one-dimensional model that captures the fundamental physics of a $p$-wave topological superconductor \cite{AKitaev_2001}. It describes spinless fermions with nearest-neighbor hopping and $p$-wave superconducting pairing. This model provides a minimal and exactly solvable framework to explore topological superconductivity. This has inspired various experimental realizations in engineered quantum systems \cite{Yao_2022_Topocircuit, PhysRevB.103.045428,Stanescu_2013,PhysRevB.107.035440, PhysRevB.98.155314}. Despite growing interest in topological materials, very few studies have explored HHG in topological superconductors \cite{Pattanayak2022,Baldelli2022},  which motivated us to investigate HHG in the spatially modulated Kitaev chain, where a site-dependent modulating potential of the strength $\lambda$ modifies the hopping amplitude. We observe distinct changes in the harmonic emission profile by tweaking the modulating potential, which offer new possibilities for XUV radiation applications \cite{Dudovich2006,Yao2011,Bengtsson_2019,Yong2017}.

The structure of the paper is as follows: In Sec. \ref{sec2}, we introduce the model and the theoretical framework used to calculate the HHG spectrum. Section \ref{sec3} presents an analysis of the system under different modulating potential scenarios. We begin by examining the effect of tuning the modulating potential in the presence of MGSs in Sec. \ref{sec3a}, followed by a dynamical excitation analysis of the HHG spectrum in Section \ref{sec3b}. In Sec. \ref{sec3c}, we discuss the role of intermediate states, and finally, we conclude the study and discuss some future directions in Sec. \ref{sec4}.
 
\section{Methodology}
\label{sec2}

\subsection{Model Hamiltonian}
In this work, we consider a spatially modulated Kitaev chain, which is equivalent to the dimerized Kitaev chain model (DKC) studied in earlier works \cite{Shilpi_Roy_DKC, Shuchen_DKC, Roy2024_sreports_DKC}. Throughout this paper, we refer to our prototype model as a dimerized Kitaev chain of the following form:
\newpage
\begin{widetext}
\begin{equation}
\begin{split}
H_0 =  \mu \sum_{j=1}^{N} \left( c_{j,A}^\dagger c_{j,A} + c_{j,B}^\dagger c_{j,B} \right) - \sum_{j=1}^{N-1} \Bigg\{ \Big[ w(1 - \lambda) c_{j,A}^\dagger c_{j,B}  + w(1 + \lambda) c_{j,B}^\dagger c_{j+1,A} \Big] - \Delta \Big[  c_{j,A}^\dagger c_{j,B}^\dagger + c_{j,B}^\dagger c_{j+1,A}^\dagger \Big] + \text{H.c.} \Bigg\},
\end{split}
\label{H_0}
\end{equation}
where $c_{j,A}^\dagger$ (or $c_{j,B}^\dagger$) is a fermionic creation operator on the sublattice A (or B) of the $j$-th unit cell, $w$ is the hopping potential (or the tunneling potential), $\mu$ is the chemical potential, and $\Delta$ denotes the $p$-wave superconducting pairing potential (or the superconducting order parameter), which is taken to be real in the present study. The dimerization parameter $\lambda$ (with $|\lambda| < 1 $) is a site-dependent variable hopping term that modulates the hopping amplitudes. The parameter $\lambda$ modulates the different hoppings as $w(1-\lambda)$ and $w(1+\lambda)$, denoting intra (inside of each sublattice) and inter (between sublattices) sublattice hoppings, respectively. We restrict the pairing potentials to be the same for both cases to work on a minimised topological model. The Hamiltonian is expanded using a basis of the structure:
\be
\Psi = \left( c_{\rm 1,A}^\dagger c_{\rm 1,A} c_{\rm 1,B}^\dagger  c_{\rm 1,B} c_{\rm 2,A}^\dagger c_{\rm 2,A} c_{\rm 2,B}^\dagger c_{\rm 2,B} \dots c_{\rm N,A}^\dagger  c_{\rm N,A} c_{\rm N,B}^\dagger c_{\rm N,B} \right)^T 
\ee
is a $2N$ dimensional vector, leading to the BdG Hamiltonian in the matrix form $H_0 = \Psi^\dagger \mathbb{H} \Psi$. The kernel Hamiltonian $\mathbb{H}$ is constructed as :
\be
\mathbb{H} = \begin{pmatrix}
\epsilon & \chi_{\rm intra} & \chi_{\rm inter} & 0 & \cdots &  \cdots &\cdots & 0 \\
\chi_{\rm intra}^\dagger & \epsilon & \chi_{\rm intra} & \chi_{\rm inter} & \cdots & \cdots & \cdots & 0 \\
\chi_{\rm inter}^\dagger & \chi_{\rm intra}^\dagger & \epsilon & \chi_{\rm intra}  & \cdots &  \cdots & \cdots & 0 \\
0 & \chi_{\rm inter}^\dagger & \chi_{\rm intra}^\dagger & \epsilon & \cdots & \cdots & \cdots & 0 \\
\vdots & \vdots & \vdots & \vdots & \ddots & \vdots & \vdots & \vdots\\
0 & 0 & 0 & 0 & \cdots & \epsilon & \chi_{\rm intra} & \chi_{\rm inter} \\
0 & 0 & 0 & 0 & \cdots & \chi_{\rm intra}^\dagger & \epsilon & \chi_{\rm intra} \\
0 & 0 & 0 & 0 & \cdots & \chi_{\rm inter}^\dagger & \chi_{\rm intra}^\dagger & \epsilon
\end{pmatrix},
\ee
where $\epsilon = \mu \sigma_z,\, \chi_{\rm intra} = -w(1-\lambda) \sigma_z + i \Delta \sigma_y,$ and $\chi_{\rm inter} = -w(1+\lambda) \sigma_z + i \Delta \sigma_y.$
\end{widetext}

The energy spectrum of the model is shown in Fig. \ref{hhg_full}(a). It exhibits a four-band structure consisting of one valence band (VB) and three conduction bands (CB1, CB2, and CB3). Intermediate states appear between each pair of bands. Based on their spectral position and localization properties, we classify these states as follows: (i) the state between the VB and CB1 is referred to as the negative mid-gap state (nMGS); (ii) the state between CB1 and CB2 is identified as a Majorana bound state (MBS); and (iii) the state between CB2 and CB3 is the positive mid-gap state (pMGS). The mid-gap states (MGSs) are non-self-conjugate edge states that appear within the energy gap of topological superconductors \cite{Bernevig2022-zx}. These states do not necessarily reside at zero energy but typically arise near quantum vortices or inhomogeneities, such as dimerization or local defects that modulate the chain's hopping potential. The MGSs are generally not topologically protected, hence these are sensitive to local variations \cite{Srinidhi_25}. In contrast, Majorana-bound states (MBS) are self-conjugate zero-energy edge states that emerge from the system’s non-trivial topological order and particle-hole symmetry \cite{Kitaev, AKitaev_2001}. Unlike MGS, the MBS are robust against local perturbations and are characterized by a non-trivial topological invariant, such as the winding number \cite{Srinidhi_25}. These states are localized at the ends of the chain and are a hallmark of the system's topological phase. This research focuses on how these intermediate states influence the harmonic spectrum. Here, the model under consideration preserves all three fundamental symmetries (time-reversal, particle-hole, and chiral). Hence, it is classified within the BDI symmetry class in the periodic table of topological materials \cite{NSSH,Wieder2021-rr, Ryu_2010}.
\subsection{Coupling to an external field}

We study the response of a strong laser field in the model. In velocity gauge \cite{Bauer2019,Chuan24}, coupling between the AAH chain and the strong field acquires a time-dependent phase factor in the hopping term by Peierls substitution as $\Exp{-iaA(t)}$, where $a$ is the lattice constant and $A(t)$ is the time-dependent vector potential that describes the shape of the laser pulse,
\be
A(t) = A_0 \sin(\omega_0 t) \exp \left[-4 \ln 2 \left( \frac{t-2T}{T} \right)^2 \right],
\ee
where the fundamental frequency of the laser field $\omega_0 = 0.00632$ a.u., and the full width at half maximum (FWHM) pulse duration $T = 1.25 \tau$ with the total time-duration considered to be $4T$. Here, $\tau$ denotes one optical cycle. The amplitude of the vector potential is $A_0=0.1$ a.u.

The time-dependent Hamiltonian is expressed as:
\begin{widetext}
\be
H(t) = \sum_{j} \Bigl\{\mu \left( c_{j,A}^\dagger c_{j,A} + c_{j,B}^\dagger c_{j,B} \right) - \Bigl[w(1 - \lambda)\Exp{-iaA(t)} c_{j,A}^\dagger c_{j,B}  + w(1 + \lambda)\Exp{-iaA(t)} c_{j,B}^\dagger c_{j+1,A}  -\Delta \Big(  c_{j,A}^\dagger c_{j,B}^\dagger + c_{j,B}^\dagger c_{j+1,A}^\dagger \Big) + \text{H.c.} \Bigr] \Bigr\},
\ee
Here, the $p$-wave pairing term is taken as a static parameter. This assumption is only justified if the superconductor is shielded from the incoming light \cite{Baldelli2022}. With the initial state prepared to be $\psi_m (t=0) = \phi_m $ ($m = 1,2,...M_{occ}$), where $\phi_m$ is the eigenstates of the field-free Hamiltonian [$H_0$ in the Eqn. \eqref{H_0}], the time-dependent wavefunction ($\psi_m(t)$) are numerically solved by using the Crank-Nicolson method \cite{Nivash_23}. Thus, the total current is computed as:
\be
J_{\rm tot}(t) = \sum_m \bra{\psi_m} \mathfrak{J}(t) \ket{\psi_m}
\ee
where $\mathfrak{J}(t)$ current operator is expressed as 
\be
\mathfrak{J}(t) = (-ia) \sum_{j} \Bigl[  w(1 - \lambda)\Exp{-iaA(t)} c_{j,A}^\dagger c_{j,B} + w(1 + \lambda)\Exp{-iaA(t)} c_{j,B}^\dagger c_{j+1,A}  - \text{H.c.} \Bigr].
\ee
\end{widetext}
The HHG spectral intensity is proportional to the absolute square of the Fourier transform of the current:
\be
S_\text{\rm tot}(\omega) = \Big| \int J_\text{\rm tot}(t) \exp(-i\omega t) dt  \Big|^2
\ee

\section{Results and Discussion}
\label{sec3}

\subsection{Band Topology and HHG Spectrum}
\label{BTP}

\begin{figure*}
\centering\includegraphics[width=1.0\textwidth]{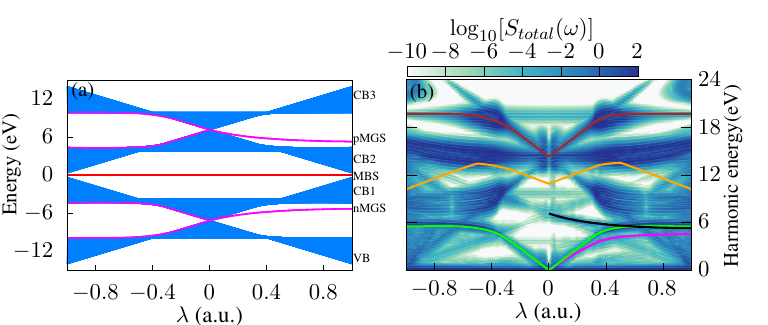}
\caption{The figure depicts the eigen spectrum (a) and the harmonic spectrum (b) for varying modulating potential $\lambda$. Other system parameters are set at $\mu = 0.225$ a.u., $w = 0.25$ a.u., $\Delta = 0.2375$ a.u. Panel (a) presents the energy spectrum of the model in Eq. \eqref{H_0}. The bulk bands are represented by blue, and the intermediate states by pink and red for MGS and MBS, respectively. Panel (b) shows the harmonic spectra corresponding to the band in panel (a). The color codes are defined in Table \ref{table}. The color bar represents the harmonic intensity.}
\label{hhg_full}
\end{figure*}

Figure \ref{hhg_full}(a) illustrates the band structure of the system as a function of the modulation parameter $\lambda$. It is evident that the band structure is symmetric about $\lambda = 0$, with one notable exception: in $\lambda \leq 0$ regime, the MGSs hybridize with the bulk states. As a result, we initially focus our analysis on the parameter regime $\lambda \geq 0$, where the MGSs remain isolated from the bulk. The implications of MGS–bulk mixing for $\lambda < 0$ will be addressed in later sections.

To better characterize the system's response, we divide the modulation parameter space into three distinct segments based on the relative positioning of the MGS within the band structure: (i) Segment I: $0 < \abs{\lambda} \leq 0.25$; (ii) Segment II: $0.26 < \abs{\lambda} \leq 0.59$; (iii) Segment III: $0.6 < \abs{\lambda} \leq 1$

The harmonic spectrum for varying $\lambda$ is shown in Fig. \ref{hhg_full}(b). Distinct pathways for harmonic enhancement are identified using different line styles, which are summarized in Table \ref{table}
\begin{table}[b]
\caption{Color code used to represent different electronic transitions contributing to the harmonic spectra. This standardized scheme is used throughout this work.}
\label{table}
\begin{ruledtabular}
\begin{tabular}{lcc}
Line color & Initial state & Final state \\ \hline
Magenta & Last of VB & nMGS \\
Green & Last of VB & First of CB1 \\
Black & nMGS & MBS \\
Orange & Last of VB & First of CB2 \\
Brown & Last of VB & First of CB3 \\
\end{tabular}
\end{ruledtabular}
\end{table}
We can make following observations from Fig. \ref{hhg_full}(b):
\begin{enumerate}
   \item  Segment I: the harmonic spectrum exhibits a substantial enhancement near the brown line, indicating a dominant transition between VB and CB3. This behavior reflects a pure Cooper-pair tunneling mechanism, where the quasiparticle predominantly occupies a chiral partner state.
   \item Segment II: multiple enhancements with equiprobable intensities are observed. This indicates the presence of several intermediate states between the bands, sufficiently separated from the bulk, enabling the quasiparticle to access all unoccupied CBs with almost equal probability. 
   \item Segment III, the bulk CBs (CB1 and CB2) approach closer to the MBS. This proximity results in a plateau-like broader enhancement across the spectrum, particularly from the magenta to the orange line region, signifying a broader range of active transitions.
\end{enumerate}

\begin{figure}
\centering\includegraphics[width=1\columnwidth]{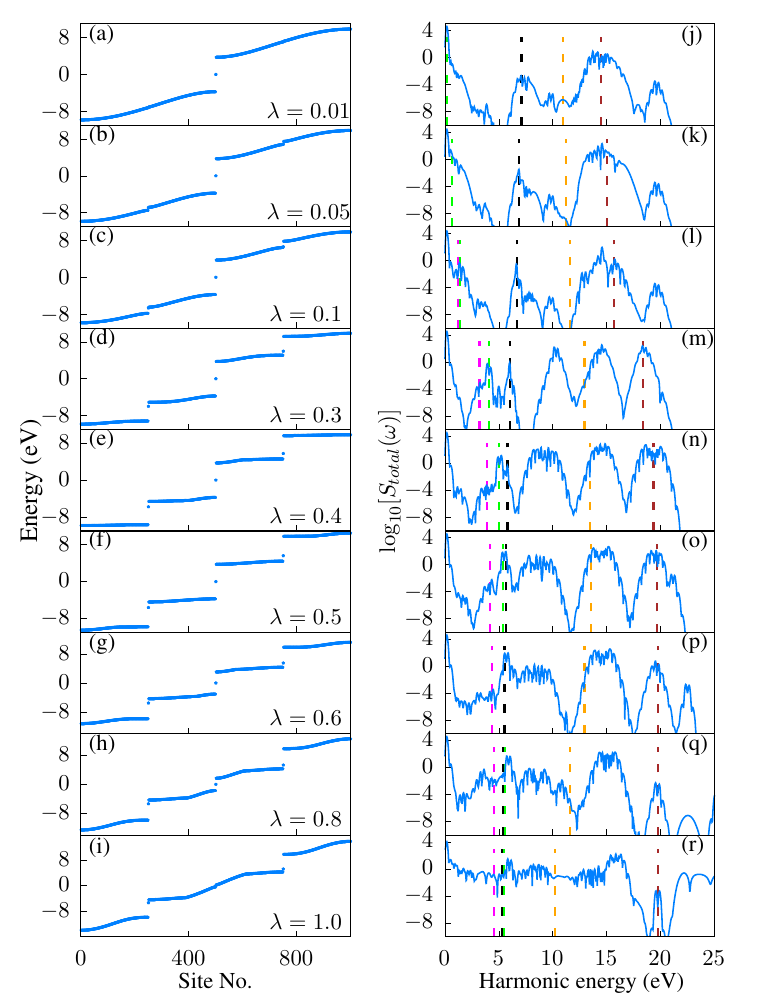}
\caption{The band-structures for different values of the parameter $\lambda$ are presented (left column) along with the associated harmonic spectra (right column). The values of the respective $\lambda$ are shown on the respective band structure. It should be noted that the band-structure (a-c) belongs to Segment I, (d-f) belongs to Segment II, and (g-i) belongs to Segment III. The color codes of each emission are discussed in the Table \ref{table}. }
\label{hhg_withMGS}
\end{figure}

\subsection{Harmonic enhancements in different segments}
\label{sec3a}

We elaborate on the results presented in Fig. \ref{hhg_full} by adjusting the modulating parameter $\lambda$ across different segments and examining the enhancements of the HHG in each segment. In Segment I, we consider modulating parameter values $\lambda = 0.01, 0.05,$ and $0.1$ a.u. The corresponding energy diagrams [Fig. \ref{hhg_withMGS}(a-c)] and harmonic spectra [Fig. \ref{hhg_withMGS}(j-l)] are presented. In this segment, the bands begin to split into four, where $\lambda$ separates the bands VB and CB1, followed by the bands CB2 and CB3. All these bands have a small finite gap in Segment I. At this point, a harmonic enhancement is observed in Fig. \ref{hhg_withMGS}(j), marked by a magenta line, followed by a plateau corresponding to the interband transition from MGS $\leftrightarrow$ MBS. No significant enhancement in the HHG is observed in the region from VB $\leftrightarrow$ CB1, due to the small gap between VB and CB1. However, the second plateau (marked by a brown line) shows a quasiparticle transition from VB $\leftrightarrow$ CB3 in the harmonic spectrum, which surpasses the first plateau. The quasiparticle surpassing the CBs(CB1 $\&$ CB2) is due to the nature of the quasiparticle basis construction, where there is a corresponding anti-particle relaxation for each quasiparticle excitation. Upon increasing $\lambda$ to $0.05$, we observe a harmonic spectrum in Fig. \ref{hhg_withMGS}(k) similar to that of the previous case [Fig. \ref{hhg_withMGS}(j)]. Moreover, when the modulating parameter $\lambda$ is increased to $0.1$, we clearly distinguish four bands with a finite bandgap [Fig. \ref{hhg_withMGS}(c)]. Subsequently, the harmonic spectrum is enhanced as observed in Fig. \ref{hhg_withMGS}(l) compared to the previous $\lambda$ values. Despite the clear separation of the bulk bands, since the MGS did not isolate from the bulk completely, we continue not to observe harmonic enhancement around the MGS region in this initial segment.

In Segment II, we consider $\lambda = 0.3, 0.4,$ and $0.5$ a.u. The harmonic spectrum has a peculiar behavior with multiple enhanced plateaus and cutoffs in Fig. \ref{hhg_withMGS}(m-o). To verify the behavior of the enhancement, we analyze different enhanced plateaus correlating them with the corresponding energy spectrum in Fig. \ref{hhg_withMGS}(d-f). The bands split equidistantly, and intermediate states are isolated from the bulk for different $\lambda$ [Fig. \ref{hhg_withMGS}(d-f)]. The interplay between the two scenarios modifies the HHG dynamics, facilitating a new VB$\leftrightarrow$MGS interband transition pathway, as shown in Fig. \ref{hhg_withMGS}(m-o). Additionally, a harmonic peak near the green line (VB $\leftrightarrow$ CB1). Neither of these harmonic enhancements occurred in Segment I. A peak near the black line (MGS $\leftrightarrow$ MBS) is observed, similar to Segment I.  The first plateau arises between the black and orange line due to interband coupling from the VB $\leftrightarrow$ MBS, as shown in Fig. \ref{hhg_withMGS}(m). When $\lambda=0.4, 0.5$, we noticed earlier enhancements in the first plateau responsible for coupling from the last CB1 to the beginning of CB2. The bandwidth of the second plateau arises from VB $\leftrightarrow$ beginning of CB2 (orange line), and the plateau terminates at VB $\leftrightarrow$ last CB2. Similarly, the bandwidth of the final plateau is predominantly governed by VB $\leftrightarrow$ CB3, which confirms quasiparticle behaviour persists in the system. Besides these noticeable observations, the bandwidth of the multiple plateaus stems from the interband coupling between the initially filled band (VB) and higher lying conduction bands via intermediate states (MGS/MBS).
\begin{figure*}
\centering\includegraphics[width=2.2\columnwidth]{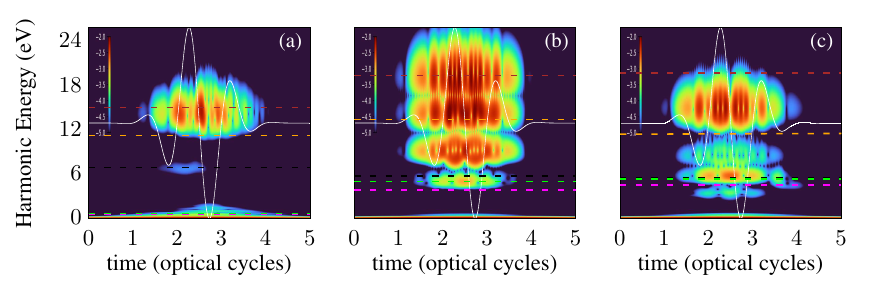}
\caption{The figure demonstrates the dynamical emission profile of the model. We have computed the time-frequency analyzed emission for three different modulating parameters: $\lambda = 0.05$ a.u. (a), $0.4$ a.u. (b), and $0.8$ a.u. (c). Other system parameters are kept the same, like the previous figures.}
\label{TF_analysis}
\end{figure*}

In the final segment, we choose $\lambda = 0.6$, $0.8$, and $1.0$ a.u., as presented in Fig.\ref{hhg_withMGS}(p-r). We observed earlier harmonic enhancement near the vicinity of the magenta line. The energy spectrum shows that the CB1 and CB2 begin to close [refer to Fig. \ref{hhg_withMGS}(g-i)]. For this scenario, a similar harmonic emission is obtained from the green and black lines as shown in Fig. \ref{hhg_withMGS}(p-r). These first and second plateaus are merged due to the high transition probability between the bulk bands CB1 and CB2. This results in a considerable bandwidth plateau from the black line to the orange line for $\lambda=0.6, 0.8$ a.u.. Moreover, the broader plateau is obtained due to bulk bands approaching near the zero-energy for the high modulating potential ($\lambda=1$). We noticed less harmonic intensity around the brown line for these three modulating potentials due to a larger bandgap.

To summarise, the harmonic emission can be effectively tuned by the modulation parameter ($\lambda$), which alters the band topology of the system. This results in distinct enhanced plateau structures: a well-defined plateau in two separate regions (Segment I), multiple plateaus with nearly equal enhancement (Segment II), and a broader bandwidth plateau with more active transitions (Segment III).

\subsection{Time frequency analysis of HHG}
\label{sec3b}

To further dig into the temporal dynamics of HHG and identify the origin of various emission features, we perform a Gabor transform and present the results obtained as a time-frequency analyzed harmonic spectrum in Fig. \ref{TF_analysis} for chosen values of modulating parameter ($\lambda=0.05,0.4,0.8$ a.u.). Figure \ref{TF_analysis} demonstrates the harmonic enhancement for a specific instant of time. The enhancement pathways denoted with color lines are similar to the previous sections. In general, the harmonic emission can be understood by the three-step procedure as follows: (i) Excitation of a quasiparticle from the occupied VB to unoccupied CB (higher-energy level) due to the applied laser field, (ii) Intra-band dynamics of the excited quasiparticle in the higher energy conduction bands, and (iii) the relaxation of the quasiparticle back to VB after reversing the polarity of the laser field. In this model, the harmonic enhancements are distinct concerning the modulating parameter characterizing the band topology. To mark this distinct behavior, we begin by setting $\lambda=0.05$. This is where the MGS is not completely isolated from the bulk. Consequently, no harmonic emission in Fig. \ref{TF_analysis}(a) around the magenta and green line. We find that the most prominent emission observed around the brown line ($\sim 14.98$ eV) corresponds to interband transitions from the VB $\leftrightarrow$ CB3 [shown in Fig. \ref{TF_analysis}(a)]. Hereby, confirming the quasiparticle nature of this system and correlates with the highly intense plateau near the brown line in the harmonic spectrum [Fig. \ref{hhg_withMGS}(k)]. The high frequency harmonics are emitted around $E(t) \sim 0$, i.e, where $A(t)$ is maximum. The less prominent emission seen near the black line ($\sim 6.87$ eV) in Fig. \ref{TF_analysis}(a), reassuring the less intense plateau around the black line in the harmonic spectrum [Fig. \ref{hhg_withMGS}(k)]. 

Proceeding to the next segment, we set $\lambda = 0.4$  (see Fig. \ref{TF_analysis}(b)). At this value, we observe harmonic emission prominently near the magenta line (VB $\leftrightarrow$ MBS), which exhibits a regular pattern closely following the laser field, indicative of a dominant single transition channel. In addition, weaker harmonic emissions are observed near the green and black lines in Fig. \ref{TF_analysis}(b). A broader range of harmonic emission emerges between the black and orange lines (around $\sim 9.6$ eV), extending beyond the orange line ($\sim 13.4$ eV) and continuing to the brown line ($\sim 19.32$ eV). This progression corresponds to the multiple plateaus observed in the harmonic spectrum [Fig. \ref{hhg_withMGS}(n)]. The harmonic emission is enhanced beyond $\sim 1.5 \tau$, which is trivial, since the driving laser pulse attains a finite strength beyond $\sim 1.5 \tau$. One has to note that, for this modulating potential, the intermediate states are away from the bulk bands. Consequently, these intermediate states provide additional pathways for a quasiparticle to climb from the occupied VB to higher unoccupied CBs. The presence of multiple transition channels is attributed to quantum interference among these pathways. Subsequently, the presence of multiple pathways attributes to the emission, which does not follow the laser profile. From literature \cite{Chaun2019,Pattanayak2020,Shao2023,Zhang2024}, we comment that this phenomenon is purely due to multiple intermediate states. 

Finally, to look at Segment III, we set $\lambda=0.8$  [Fig. \ref{TF_analysis}(c)]. Mild emission is observed around the magenta line, which is followed by the green and black lines, exhibiting dynamics obeying the driving field. The broader bandwidth plateau in the harmonic spectrum in Fig. \ref{hhg_withMGS}(q) correlates with the harmonic emission emerging from the magenta line and continuing till the orange line. This is attributed to the band-closing (CB1 and CB2). Further, in addition to the plateau observed, the high-intensity peak observed in between the orange and the brown line in Fig. \ref{hhg_withMGS}(q), contributes to the high-intensity emission seen around the same region in Fig. \ref{TF_analysis}(c). This reassures the phenomenon that multiple pathways lead to an interference pattern, where the emission is irrespective of the laser profile.
\begin{figure}
\centering\includegraphics[width=1\columnwidth]{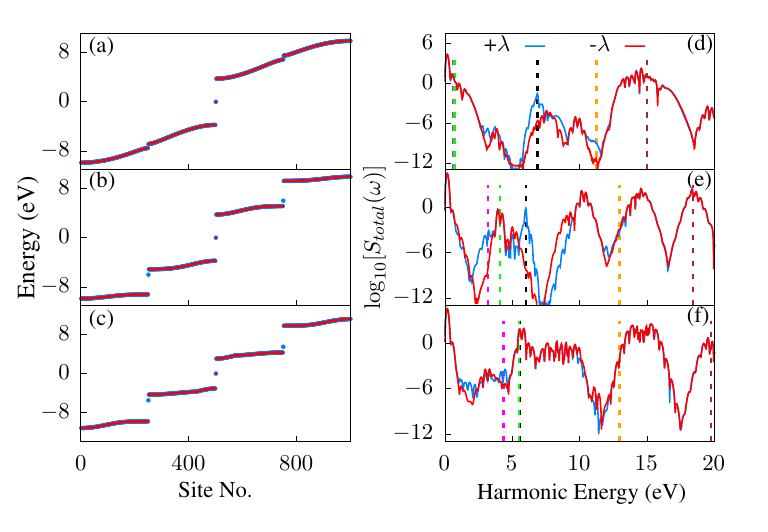}
\caption{To investigate the role of intermediate states, we compute the harmonic spectra for chosen modulating potentials: (a) $\abs{\lambda}=0.05$ a.u.; (b) $\abs{\lambda}=0.3$ a.u.; (c) $\abs{\lambda}=0.6$ a.u. The red denotes the hybridised MGS cases, whereas the blue denotes the isolated MGS cases for varying $\lambda$. }
\label{hhg_compare}
\end{figure}

\subsection{Role of the intermediate states}
\label{sec3c}

From the previous sections, it has been established that intermediate states create additional pathways for quasiparticles, leading to multiple equi-probable emissions in the harmonic spectrum. To further explore the role of these intermediate states, we extend our analysis by lifting the degeneracy of one of these doubly degenerate states. To isolate this effect, we focus on the negative side of the modulating potential ($\lambda \leq 0$), where the MGS hybridize with the bulk bands, as illustrated in Fig. \ref{hhg_full}(a). This can be interpreted as the system being in a topological state, with only the Majorana-bound states (MBS) remaining localized. In the subsequent analysis, we examine various modulating potentials corresponding to isolated MGS ($\lambda=0.05, 0.3, 0.6$ a.u.) and hybridized MGS ($\lambda=-0.05, -0.3, -0.6$ a.u.), as shown in Fig. \ref{hhg_compare}. The red line in the figure represents the case of hybridized MGS ($-\lambda$), while the blue line corresponds to isolated MGS ($+\lambda$).

Starting with $\abs{\lambda} = 0.05$ (Fig. \ref{hhg_compare}(a,d)), no significant enhancement is observed near the magenta line, since the MGS is not sufficiently separated from the bulk bands, leading to similar behavior in both cases. However, discrepancies appear between the green and orange lines. The enhancement seen at the black line ($\sim 6.87$ eV) for $\lambda=0.05$ a.u. (where no enhancement occurs for $\lambda=-0.05$ ) can be attributed to quasiparticle excitation between the MGS and MBS, which is absent in the $\lambda=-0.05$ case due to the isolation of the MGS from the bulk bands. On the other hand, the enhancement near the green line ($\sim 7.49$ eV) for $\lambda=-0.05$ results from bulk transitions (VB $\leftrightarrow$ CB1) as no intermediate states are available. Beyond this region, no significant changes are observed in the harmonic spectra, indicating that quasiparticle behavior remains unaffected even when the MGS hybridizes with the bulk.

Moving to Segment II, where $\abs{\lambda} = 0.3$ (Fig. \ref{hhg_compare}(b,e)), the MGS is now sufficiently distanced from the bulk bands, leading to a harmonic enhancement around the magenta line ($\sim 3.19$ eV), corresponding to a transition from VB $\leftrightarrow$ MGS. A subsequent enhancement near the black line ($\sim 5.99$ eV) suggests quasiparticle excitation between MGS and MBS. These transitions are forbidden for $\lambda=-0.3$ due to the hybridization of the MGS, and the harmonic enhancement observed near the green line ($\sim 4$ eV) corresponds to bulk transitions (VB $\leftrightarrow$ CB1) in both cases.

Further, in Segment III, the emission patterns for $\lambda=0.6$ and $\lambda=-0.6$ are similar, with a slight difference observed near the magenta line (VB $\leftrightarrow$ MGS). The bulk bands CB1 and CB2 approach the zero-energy state for this modulating potential. As a result, transitions from MGS $\leftrightarrow$ MBS (black line) and from VB $\leftrightarrow$ CB1 (green line) produce similar harmonic emissions, causing the two lines to overlap, as shown in Fig. \ref{hhg_compare}(f).

Finally, this study highlights the significant role of intermediate states in the harmonic spectra, particularly through the modulation of MGSs and their interaction with bulk bands. Isolating MGS leads to distinct harmonic enhancements due to quasiparticle excitations between MGS and MBS. At the same time, hybridized MGS exhibits similar behavior exclusively for bulk transitions, as the other transition is forbidden. Although this approach could be extended to the trivial phase, resulting in trivial bulk transitions, we focus on the topological phase to better distinguish the impact of intermediate states, whether isolated or hybridized.

\section{Summary}
\label{sec4}
In this work, we have investigated high-harmonic generation (HHG) in a dimerized Kitaev chain, where a site-dependent modulating parameter $\lambda$ influences the hopping amplitudes and decides the system's topology. By varying $\lambda$ within the range $[-1.0, 1.0]$, we identified distinct harmonic emission profiles, which we classified into three segments. In Segment I, we have observed two plateau structures, with the dominant plateau corresponding to quasiparticle transitions to the chiral partner state, confirming the quasiparticle nature of excitations in the topological superconducting (TSC) phase. Segment II has displayed a multiple plateau structure, where intermediate states facilitated equiprobable transitions to higher conduction bands, creating multiple emission pathways. In Segment III, broader plateaus have emerged, reflecting active transitions between the bands. The time-frequency analysis further corroborated the diversity of harmonic emissions across these segments.

Furthermore, by tuning $\lambda \leq 0$, we induced the hybridization of the MGS with the bulk, suppressing earlier harmonic enhancements, like the enhancements due to intermediate states (near the magenta, green lines). This highlights the critical role of intermediate states, particularly when MGSs are isolated. Our findings demonstrate that the harmonic emission profile can be selectively controlled by adjusting $\lambda$, providing a method for tailoring HHG in topologically nontrivial systems. This work opens new avenues for manipulating HHG based on the topological properties of the system, with potential applications in quantum materials and topological photonics.

\section{Acknowledgments} 
The authors acknowledge the Department of Science and Technology (DST) for providing computational resources through the FIST program (Project No. SR/FST/PS-1/2017/30).

\bibliography{Ref}

\end{document}